\documentclass
[preprintnumbers,preprint,tightenlines,prd,amsmath,amssymb,nofootinbib]
{revtex4-1}

\usepackage{xspace}
\usepackage[utf8]{inputenc}
\usepackage{graphicx}

\usepackage{color}

\newcommand{\order}{\mathop{\mathcal{O}}\nolimits}
\newcommand{\meV}{\mathrm{meV}}

\newcommand{\GeV}{\mathrm{GeV}}

\newcommand{\phif}{\phi_+}
\newcommand{\phit}{\phi_-}
\newcommand{\phift}{\phi_\pm}
\newcommand{\MPl}{M_\text{Pl}}
\newcommand{\FE}{V_\text{np}}
\newcommand{\Vpert}{V_\text{pert}}
\newcommand{\Vnloop}[1]{V_{#1}}

\begin{document}

\preprint{KIAS--Q19001}
\title{Higgs potential near-criticality from a de Sitter swampland-like condition}
\author{Jae-hyeon~Park}
\affiliation{Quantum Universe Center,
  Korea Institute for Advanced Study,
  85 Hoegiro Dongdaemungu,
  Seoul 02455, Republic of Korea}

\begin{abstract}
  Motivated by the recently proposed de Sitter swampland conjecture,
  a formally same condition imposed instead
  on the \emph{convex} and \emph{real} exact effective potential
  is contemplated.
  Compared to the original conjecture,
  the modified condition admits a broader class of
  low-energy effective theories
  such as those with local maxima and/or
  false de Sitter vacua with some restrictions,
  as long as there is an anti--de Sitter vacuum.
  The observed accelerating expansion of the universe might therefore be
  attributed to a quintessence or a metastable vacuum.
  The former solution can be simplified and thus is
  better compatible with phenomenological constraints
  thanks to the convexity of the effective potential.
  Among the latter class of solutions
  is found the enthralling possibility that
  the modified condition is in fact behind the
  experimentally favoured metastability of
  the Higgs potential with an instability scale
  below or around the Planck scale.
\end{abstract}
\maketitle


A quantum field theory (QFT) is said to belong to the landscape,
if it is a low-energy effective field theory (EFT) of string theory.
Otherwise, it is said to belong to the swampland
\cite{Vafa:2005ui,Ooguri:2006in}.
These classifications were introduced as
certain classes of theories could not be related to string theory
despite enormous efforts.
Motivated by such theoretical experience,
criteria have been conjectured
that are supposed to characterize EFTs in the landscape or swampland,
see for a review \cite{Brennan:2017rbf}.

In particular,
the lack of any known rigorous construction of
a de Sitter (dS) vacuum from string theory
has recently led to
the original version of the dS swampland conjecture
\cite{Obied:2018sgi},
\begin{equation}
  \label{eq:old dS swampland conjecture}
    \MPl\, | \nabla V | > c\, V,
    \quad 0 < c \sim \order(1)
    ,
\end{equation}
where
$\MPl$ is the
reduced Planck mass,
$V$ is the scalar potential in the EFT,
$\nabla V$ is its gradient with respect to the scalar fields
with its norm defined using the metric on the field space.
Indeed,
this inequality says that $V$ at any extremum must be negative,
thereby excluding any EFT with a dS extremum from the landscape.
Still to keep our dS universe within the landscape,
we are assumed to be living not at a (false) vacuum
but on a nonvanishing slope of $V$ in the direction of
some scalar field usually dubbed a quintessence
\cite{quintessence,Tsujikawa:2013fta}.

This conjecture has subsequently been refined
to incorporate as an alternative condition the inequality
\cite{Garg:2018reu,Ooguri:2018wrx},
\begin{equation}
  \label{eq:supplementary dS swampland condition}
    \MPl^2\, \min (\nabla_i \nabla_j V) \le - c'\, V,
    \quad 0 < c' \sim \order(1)
    ,
\end{equation}
which bounds the minimum eigenvalue of the Hessian of $V$
in an orthonormal frame.
(Other ways of refinement have also been proposed
\cite{other refinements}.)
This refinement followed
as difficulties had been encountered
when the original conjecture~\eqref{eq:old dS swampland conjecture}
was applied to firmly established phenomenological particle physics models.
The root of the troubles was
that~\eqref{eq:old dS swampland conjecture}
forbids any local maximum of $V$ with a positive value
which appears necessarily in EFTs with
(spontaneous) symmetry breaking.
Known instances include the centre of the Higgs potential
\cite{
  Denef:2018etk,
  Murayama:2018lie,
  Choi:2018rze,
  Hamaguchi:2018vtv%
}
as well as
the local maxima found in the periodic potentials of the neutral pion
\cite{Choi:2018rze}
and the hypothetical
QCD axion
\cite{Murayama:2018lie}.
These counterexamples are admitted
by the refined conjecture~(\ref{eq:old dS swampland conjecture},%
\ref{eq:supplementary dS swampland condition}).
%
%
The swampland conjectures have also been put to the test
in the context of cosmic inflation,
to reveal differing degrees of compatibility with the models
\cite{applications to inflation}.


As the dS swampland conjectures restrict the scalar potential,
a brief review of its different formulations should be in order.
First of all, one can define the generating functional,
\begin{equation}
  \label{eq:partition function}
  Z[J] =
  \int\!\!\mathcal{D}\phi
  \exp
  \left\{
    i \int\!\!d^4x\, \bigl( \mathcal{L}[\phi(x)] + J(x)\phi(x) \bigr)
  \right\}
  ,
\end{equation}
in terms of the classical Lagrangian $\mathcal{L}$ and
the external currents $J$ as sources for the fields $\phi$.
The connected generating functional is then given by
\begin{equation}
  \label{eq:connected generating functional}
  W[J] = -i\, \ln Z[J]
  ,
\end{equation}
of which the Legendre transform yields
the one-particle irreducible 
effective action
\cite{JonaLasinio:1964cw},
\begin{equation}
  \label{eq:effective action}
  \Gamma[\phi] = W[J] - \int d^4 x J(x) \phi(x)
  .
\end{equation}
The effective potential can be defined
as $\Gamma[\phi]$
specialized to coordinate-independent field expectation values
\cite{Jackiw:1974cv},
\begin{equation}
    \FE(\phi) =
    \left. - \frac{1}{VT}\, \Gamma[\phi] \right|_{\phi = \text{const.}}
    ,
\end{equation}
with the spacetime volume $VT$ factored out.
This nonperturbatively defined exact effective potential
$\FE$ can be approximated by a perturbation series
in the form,
\begin{equation}
  \label{eq:free energ}
  \FE \simeq
  \Vpert \equiv
  \Vnloop{0} +
  \Vnloop{1} +
  \Vnloop{2} + \cdots
  ,
\end{equation}
where
$\Vnloop{0}$ coincides with the tree-level potential in
the classical Lagrangian $\mathcal{L}$,
and $\Vnloop{1,2,\ldots}$ are the loop corrections
at each order.
It shall be understood that $\Vpert$ may also include
nonperturbative contributions as in the potential of
an axion or mesons.

It is well known that
$\FE$ and
$\Vpert$ are gauge \cite{Jackiw:1974cv}
and renormalization scale dependent,
but their values at the extrema are not
\cite{gauge independence of potential extrema}
and are regarded as physical quantities,
see for scale dependence e.g.\
\cite{Andreassen:2014eha,Andreassen:2014gha}.
This unphysical nature of effective potentials
stems from the fact that their arguments i.e.\
the scalar field values are not physical quantities.
In perturbation theory,
the scale dependence of $\Vpert$ is reduced
as higher and higher order loops are included.
This makes it mandatory to perform
an all-order resummation of the large logarithms,
when there is an orders-of-magnitude separation
between the scale at which the input parameters are fixed
and the scale at which $\Vpert$ is to be evaluated.

Another remarkable property of $\FE$ (but not of $\Vpert$)
is its convexity
\cite{Symanzik:1969ek}.
This means especially
that $\FE$ does not have any local maximum even if
$\Vnloop{0}$ or $\Vpert$ does.
The shape of $\FE$ between the two local minima
$\phi_1$ and $\phi_2$ of $\Vpert$ is linear
\cite{Fujimoto:1982tc,free energy linearity,Cahill:1993mg},
and thus can be approximated by the linear interpolation
(see e.g.\ \cite{Weinberg:1996kr,Peskin:1995ev}),
\begin{equation}
  \label{eq:maxwell construction}
  \FE\left(x \phi_1 + (1-x) \phi_2\right) \simeq
  x \Vpert(\phi_1) + (1-x) \Vpert(\phi_2)
  ,
  \quad
  0 < x < 1
  ,
\end{equation}
analogous to
the Maxwell construction for free energies in thermodynamics,
see e.g.\ \cite{Peskin:1995ev}.
This has also been a traditional way to resolve doubts about
the imaginary part
of $\Vpert$ that develops at points where $\Vnloop{0}$ is concave
\cite{Jackiw:1974cv,Fujimoto:1982tc,Cahill:1993mg}.
The above construction is guaranteed to inherit
the reality of $\FE$ originating from its definition.

%
%
Recalling these properties of effective potentials,
one might then ask a natural question:
which $V$ does the dS swampland conjecture concern?
%
%
In this work,
the quantum limit of choosing $\FE$ shall be entertained,
in which case
the original conjecture~\eqref{eq:old dS swampland conjecture}
would read
\begin{equation}
  \label{eq:free energy condition}
    \MPl\, | \nabla \FE | > c\, \FE,
    \quad 0 < c \sim \order(1)
    .
\end{equation}
As has been pointed out recently \cite{Kobakhidze:2019ppv},
this modification weakens~\eqref{eq:old dS swampland conjecture}
to such an extent that
the supplementary condition~\eqref{eq:supplementary dS swampland condition}
would not be needed to accommodate
the particle physics models with a local maximum
in the potential
\cite{
  Denef:2018etk,
  Murayama:2018lie,
  Choi:2018rze,
  Hamaguchi:2018vtv%
}.  
Paraphrased in terms of $\Vpert$,
\eqref{eq:free energy condition} does not prohibit
any of its local maxima, dS or not, since they are all flattened
in $\FE$.
For the same reason, it is obviously more permissive than
the original~\eqref{eq:old dS swampland conjecture}
as well as the refined dS swampland
conjectures~(\ref{eq:old dS swampland conjecture},%
\ref{eq:supplementary dS swampland condition}).
Note that~\eqref{eq:supplementary dS swampland condition}
becomes redundant, thereby rendering
both versions of the conjectures equivalent,
if $V$ therein is replaced by $\FE$
\cite{Kobakhidze:2019ppv}.
A global minimum still needs to be negative.


An immediate consequence of~\eqref{eq:free energy condition}
would therefore be that
a dS space cannot be ascribed to a global minimum of $\Vpert$.
Given the astronomical observations indicating
a small but positive vacuum energy,
this would imply that we are not living in a true vacuum.
There are two ways to negate the last noun phrase:
(a) a false vacuum, or
(b) a non-vacuum.
Note that the former possibility has not been mentioned
in Ref.~\cite{Kobakhidze:2019ppv}.


Option (b) would mean that
$| \nabla \Vpert | > 0$ in fact where we are living
owing to a non-vanishing slope in the direction of an extra scalar field
such as a quintessence.
This bears a similarity to how a quintessence reconciles
the original dS conjecture~\eqref{eq:old dS swampland conjecture}
with the cosmological constant 
\cite{Obied:2018sgi}.
A major difference is however that
the local maximum of the perturbative Standard Model (SM) Higgs potential
would not require a complicated interaction between
the quintessence and the Higgs \cite{Denef:2018etk},
as $\FE$ has no local maximum.
It would therefore be enough to add quintessence terms
simply to the Higgs potential,
thereby eluding the constraints from
a long-range force and
time dependence of the proton-to-electron mass ratio
\cite{Hamaguchi:2018vtv}.

\begin{figure}
  \centering
  \includegraphics[width=.48\textwidth]{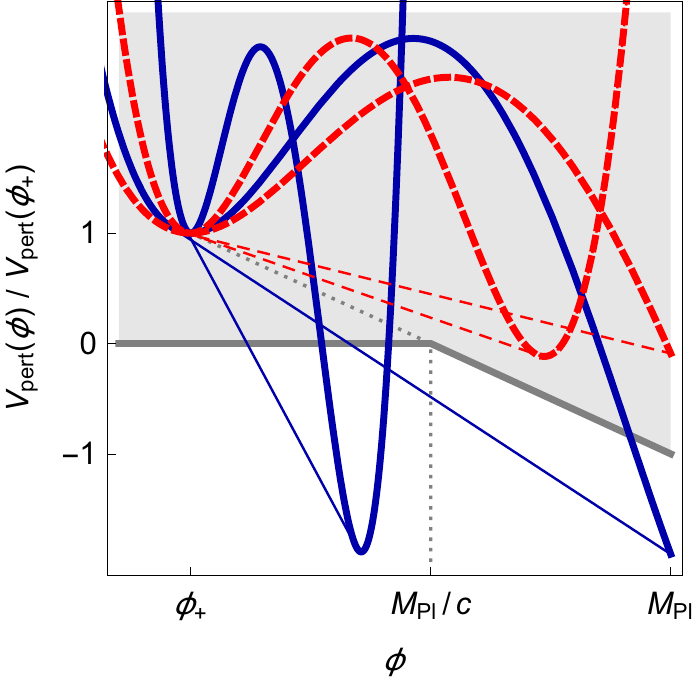}
  \hfill
  \includegraphics[width=.48\textwidth]{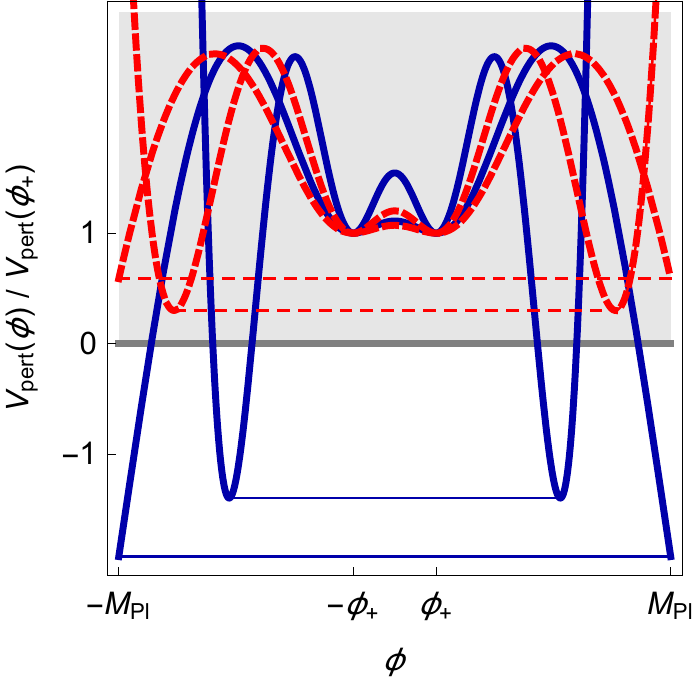}
  \caption{Schematic shapes of perturbative scalar potentials (thick)
    allowed (solid) or
    disallowed (dashed) by~\eqref{eq:free energy conditions}
    and the corresponding ``Maxwell constructions'' (thin),
    for $c > 1$.
    The global minima of each potential cannot lie in the light grey region.
    For $c < 1$,
    the horizontal part of the grey line in the left panel would extend
    up to the right border.}
  \label{fig:maxwell}
\end{figure}

With option (a),
one can tell further details of the metastability
with the aid of a ``Maxwell construction''.
First, imagine a perturbative potential having only two minima
which are connected by the scalar field direction $\phi$.
Replacing $\FE$ between these two minima
by~\eqref{eq:maxwell construction} 
lets~\eqref{eq:free energy condition}
be represented by a pair of inequalities,
\begin{subequations}
  \label{eq:free energy conditions}
  \begin{align}
    \label{eq:free energy slope}
    \Vpert(\phif) &<
    \frac{\MPl}{c}
    \left|
      \frac{\Vpert(\phif) - \Vpert(\phit)}{\phif - \phit}
    \right|
    ,
    \\
    \label{eq:free energy minimum}
    \Vpert(\phit) &< 0
    ,
  \end{align}
\end{subequations}
where $\phift$ are the field values at
our false and the true vacua, respectively.
These conditions are depicted in the left panel of Fig.~\ref{fig:maxwell},
where $c$ is assumed to be larger than unity
as suggested by some reported lower bounds on $| \nabla V | / V$
in type IIA/B compactifications
\cite{Obied:2018sgi}.
Essentially,
the global minimum $\Vpert(\phit)$
is constrained to be firstly negative and then deep enough
to meet the lower bound on $| \nabla\FE(\phif) |$.
It might happen that $\Vpert(\phit)$ is
at the end of the validity range of QFT
which is assumed to be $\MPl$
based on the distance conjecture \cite{Ooguri:2006in}.
In such a case, \eqref{eq:free energy minimum} would not apply
as the slope does not vanish at $\phit$.
If the cutoff on $|\phi|$ is smaller than $\MPl/c$,
there would be another logical possibility that
$\Vpert(\phi)$ does not necessarily need to cross the grey line.
It would only need to cross the slanted grey dotted line and then
hit the cutoff at a nonvanishing slope.
%
Numerically,
$\Vpert(\phif) \sim \order(\meV^4)$ would be in most cases negligible
compared to typical scales characterizing $\Vpert$
such as the width and height of its barrier.
Therefore, the grey upper bound on $\Vpert(\phit)$ with a mild kink
would be approximated excellently by a straight horizontal line in
the left panel of Fig.~\ref{fig:maxwell}.

A reflection symmetry of the potential would simplify the scrutiny
as illustrated in the right panel of Fig.~\ref{fig:maxwell}.
The ``Maxwell construction'' would then be
a constant between the two global minima
which~\eqref{eq:free energy condition}
would require to be negative.
In this case,
$\Vpert(\phi)$ would need to cross zero
within the validity range of QFT,
whether the slope vanishes at the global minima or not.

Subdividing option (a),
the metastability of our electroweak vacuum might be attributed to
(a1) a new direction in the field space, or
(a2) the SM Higgs.
%
%
Once an extra field is admitted as in case (a1),
it is straightforward to build a model with metastability
and there are already many such examples as:
supersymmetry with charge and/or colour breaking minima
\cite{Claudson:1983et,motivated CCB,Park:2018bkn},
metastable supersymmetry breaking sectors
\cite{Intriligator:2006dd},
relaxion mechanism
\cite{Graham:2015cka},
scalar extensions of the Higgs sector
\cite{Staub:2017ktc},
among others.
In models with a stable Higgs potential,
metastability would thus serve as a hint on
the vacuum structure altered by
the additional fields.


The remaining most predictive scenario would be
case (a2) without any extra field responsible for the metastability.
This might be
because there are no extra scalars at all or none of them
induces a deeper minimum.
An obvious but fascinating implication would then be that
the perturbative SM Higgs potential is destined to be metastable.

In this case,
a more concrete statement can be made
in the context of the SM\@. 
Assuming that $\phit \gg \phif \approx 246\, \GeV$,
one can use the renormalization group (RG) improved tree-level potential,
$\Vpert(\phit) = \lambda(\phit)\, \phit^4 / 4$,
with the running quartic Higgs coupling $\lambda(\mu)$
at the renormalization scale $\mu = \phit$.
The preceding discussions about condition~\eqref{eq:free energy condition}
on symmetric potentials
would then lead to the requirement that
$\lambda(\mu)$ turn negative
below or around $\MPl$,
an upper limit due to the distance conjecture
\cite{Ooguri:2006in}
as well as
the inherent ultraviolet cutoff of the EFT 
within which the $\beta$-function of $\lambda$ is computed.


It is amusing to notice that
this prediction of an instability scale below or around $\MPl$
is pleasantly consistent with the already favoured interpretation
of the experimental data,
even though it is
demanded independently of the low-energy boundary conditions
on the running couplings.
The RG evolution of $\lambda$ has been analysed
employing higher order corrections at the accuracy of
2-loop matching at the weak scale plus 3-loop running up to high scales
\cite{
  lambda running,
  Andreassen:2014gha,
  Bednyakov:2015sca%
}.
A first resulting feature to notice
is that $\lambda(\mu)$ maintains
a single minimum, positive or negative,
around $10^\text{17--18}\,\GeV$
while the top quark and Higgs masses as well as
the strong coupling constant are varied by $\pm 3 \sigma$
\cite{lambda running}.
The upper end of this range
would therefore bound a zero of $\lambda(\mu)$ from above,
if it exists.
Then, the central values of the SM parameters point to
metastability of the Higgs potential with
$\lambda(\mu)$ crossing zero
at a scale around $\mu \sim 10^{10}\, \GeV$,
which can vary between $10^8$ and $10^{18}\, \GeV$
if $3\sigma$ uncertainties in the data are taken into account.
Additional ambiguity in the instability scale
arising from its gauge dependence
has been investigated numerically
\cite{gauge dependence of instability scale}.


There have been attempts to understand
this intriguing selection of a special point in the parameter space:
there might be an underlying theory
which brings the SM to that particular point via the matching conditions
\cite{
  vanishing lambda from matching,
  Park:2018bkn%
};
near-criticality might be an attractor within the multiverse
\cite{Giudice:2006sn}.
Yet another inspiring possibility
would be that inequality~\eqref{eq:free energy condition}
is in fact the reason behind the metastability of the Higgs potential,
if the condition has relevance to
physical laws of nature
such as the still developing theory of quantum gravity.


It is hard to judge whether~\eqref{eq:free energy condition}
has something to do with string theory or not.
As it turns out,
the condition is at variance with
the original motivation for~(\ref{eq:old dS swampland conjecture},%
\ref{eq:supplementary dS swampland condition}), i.e.\
to exclude dS minima.
On the contrary,
\eqref{eq:free energy condition}
admits false dS vacua albeit with restrictions on $|\nabla\FE|/\FE$.
Neither has it been proved however that
the landscape contains no dS vacua.
%
%
From the field theoretic point of view,
the use of $\FE$ in conjecture~\eqref{eq:old dS swampland conjecture}
has been advocated
\cite{Kobakhidze:2019ppv},
emphasizing the necessity of
large scale nonperturbative scalar field fluctuations
for a consistent low-energy description of the theory.
In particular,
it is not clear how to interpret the dS criteria
as imposed on $\Vpert$ if it has an imaginary part.
This doubt would be naturally resolved by employing $\FE$ instead
which is real-valued by definition.
In any case, one might make at least the following conservative statements.
It is a natural extrapolation of~(\ref{eq:old dS swampland conjecture},%
\ref{eq:supplementary dS swampland condition})
to incorporate into $V$ therein
as many quantum effects as there are.
The criterion thus modified is less restrictive
than the original but not trivial either,
and furthermore suggests intriguing phenomenological scenarios
including a potential solution
to a big puzzle raised by accelerator physics.


To sum up,
a theoretical constraint on quantum field theoretic models
has been considered.
Its form is identical to the original dS swampland conjecture
except that the effective potential is assumed to integrate
all possible (non)perturbative quantum effects,
thereby guaranteeing its reality.
Due to the convexity of the exact effective potential $\FE$,
the modified condition is more permissive than the original
as well as the refined conjectures.
Specifically,
it accepts local maxima and
false dS vacua in the perturbative potential $\Vpert$
as long as the slope of $\FE$ is everywhere steep enough,
although the true vacua must still be anti--de Sitter.
This naturally resolves conflicts with the essential local maxima
in established particle physics models.
Moreover, it opens up the possibility of attributing
the observed positive cosmological constant
not only to a quintessence but also to a metastable dS vacuum.
In the former case,
the quintessence might avoid phenomenological obstacles
thanks to its simpler interactions with the Higgs.
In the latter case,
the Higgs potential would need to 
have a deeper minimum
unless the metastability is caused by an extra scalar field.
The Higgs instability scale is then predicted to be below or around $\MPl$,
from the distance conjecture plus
the ultraviolet cutoff of the EFT\@.
This
might shed light on
the well known
near-critical structure of the SM Higgs potential
for the preferred values of low-energy data
including the top quark and Higgs masses.

\vspace{1ex}

The author thanks
Jason Evans,
Lucien Heurtier,
Jong-Chul Park,
Seodong Shin, and
Seokhoon Yun,
for the informative and encouraging discussions.
This work has been highly boosted by
the focus research program
``Dark Matter as a Portal to New Physics''
supported by the Asia Pacific Center for Theoretical Physics.

\end{document}